\documentclass[aip,pof,onecolumn,floatfix,superscriptaddress,footinbib,reprint]{revtex4}

\usepackage[pdftex]{graphicx}
\usepackage[pdftex]{hyperref}
\usepackage{natbib}
\usepackage{placeins}
\usepackage{bm}% bold math 
\usepackage{longtable,multirow,booktabs}
\usepackage{rotating}
\usepackage{amsmath,amssymb}

\usepackage{color}
\definecolor{darkblue}{rgb}{0,0,.5}

\usepackage[cdot,squaren]{SIunits}
\newcommand{\Hz}{\;\hertz}
\newcommand{\cm}{\;\centi\metre} 
\newcommand{\mm}{\;\milli\metre} 
\newcommand{\fig}[1]{Fig.~{{\ref{#1}}}} % Fig
\newcommand{\gleich}[1]{Eq.~\protect\eqref{#1}} % Eq
\newcommand{\abc}[1]{{\small{\textbf{#1}}}}
\newcommand{\eps}{\varepsilon}
\newcommand{\acct}{\ensuremath{a_\text{rms}}}
\newcommand{\atrans }{\ensuremath{\vec{a}_\text{trans}}}
\newcommand{\aSP}{\ensuremath{\vec{a}_\text{SP}}}
\newcommand{\fimp}{\ensuremath{f_\text{imp}}}
\newcommand{\tcorr}{\ensuremath{\tau_{\text{corr}}}}
\newcommand{\karman}{von~K\'arm\'an~}
\newcommand{\Karman}{Von~K\'arm\'an~}

\providecommand{\abs}[1]{\left\lvert#1\right\rvert} 
\providecommand{\mean}[1]{\left\langle#1\right\rangle} 
\providecommand{\meanB}[1]{\big\langle#1\big\rangle} 

\providecommand{\kla}[1]{\left(#1\right)} 
\providecommand{\klaB}[1]{\big(#1\big)} 
\providecommand{\bra}[1]{\left[#1\right]}

\DeclareMathOperator{\PDF}{PDF}

\newcommand{\matrize}[1] {\smash{\underline{\underline{\mathbf{#1}}}}}
\renewcommand{\vec}{\bm}
\renewcommand{\d}{\text{d}}

\hypersetup{pdftex=true, colorlinks=true, breaklinks=true, linkcolor=darkblue, menucolor=darkblue,  urlcolor=darkblue}

\begin{document}

\title{Characterizing flows with an instrumented particle measuring Lagrangian accelerations}

\author{Robert Zimmermann}
\affiliation{Laboratoire de Physique, CNRS UMR 5672, Ecole Normale Sup\'erieure de Lyon, 46 all\'ee d'Italie, F-69007 Lyon, France}
\email{robert.zimmermann@ens-lyon.org}
\author{Lionel Fiabane}
\affiliation{Laboratoire de Physique, CNRS UMR 5672, Ecole Normale Sup\'erieure de Lyon, 46 all\'ee d'Italie, F-69007 Lyon, France}
\author{Yoann Gasteuil}
\affiliation{smartINST S.A.S., 46 all\'ee d'Italie, F-69007 Lyon, France}
\author{Romain Volk}
\affiliation{Laboratoire de Physique, CNRS UMR 5672, Ecole Normale Sup\'erieure de Lyon, 46 all\'ee d'Italie, F-69007 Lyon, France}
\author{Jean-Fran\c{c}ois Pinton}
\affiliation{Laboratoire de Physique, CNRS UMR 5672, Ecole Normale Sup\'erieure de Lyon, 46 all\'ee d'Italie, F-69007 Lyon, France}

\date{\today}

\begin{abstract}
We present in this article a novel Lagrangian measurement technique: an instrumented particle which continuously transmits the force/acceleration acting on it as it is advected in a flow. We develop  signal processing methods to extract information on the flow from the acceleration signal transmitted by the particle. Notably, we are able to characterize the force acting on the particle and to identify the presence of a permanent large-scale vortex structure.
Our technique provides  a fast, robust and efficient tool to characterize flows, and it is particularly suited to obtain Lagrangian statistics along long trajectories or in cases where optical measurement techniques are not or hardly applicable.
\end{abstract}
\maketitle

%\NOTE{We demonstrate that it is possible to build auto-correlation functions of acceleration moments that depends or not on the flow regime, depending on the order of the acceleration moment.}

Turbulence is omnipresent in nature and in industry, and has received much attention for years.
In the specific field of experimental fluid dynamics research, very significant progress has been achieved during the last decade with the advent of space and time resolved optical techniques based on high speed imaging~\cite{book:expFluids}. 
However, a direct resolution of the Eulerian flow pattern is still not always possible nor simple to carry out. 
In this context, Lagrangian techniques, in which the fields are monitored along the trajectories of particles, provide an interesting alternative~\cite{annRev:LagProps,Shraiman:2000vu} with information about the small scales of turbulence (especially isotropy) and a huge focus on the particle's Lagrangian acceleration that directly reflects the turbulent forces exerted on the particles~\cite{Voth:2002hc, Mordant:2003,qureshi2007turbulent, qureshi2008acceleration,Volk:2008,Xu:2008tc}.
%It is in particular well known that fluid particles exhibit highly non-Gaussian acceleration probability density function (PDF) with stretched tails at high acceleration events~\cite{Voth:2002hc}.

From an experimental point of view several inconveniences arise. 
 In the  Lagrangian frame one would like to collect long trajectories.  However, even in confined flows it is difficult to track even just a few particles over a long time using the existing methods. For instance to use optical methods, the flow must be entirely observed and continuously recorded, something which is not yet possible. Apart from its implication for computing converged statistical quantities, several theories such as the fluctuation theorem necessitate long trajectories instead of many short ones. 
Another issue is the possible rotation of large particles in a flow, and the influence of this possible rotation on the dynamics of the particle. An optical technique following simultaneously particle position and absolute orientation in time    has been recently developed~\cite{Zimmermann:2011uu}. It shows in particular that for increasing turbulence, solid particles experience stronger rotation~\cite{Zimmermann:2012th,liftforce}.
The technique used in those experiments is not straightforward and needs careful calibration and synchronization, as well as an expensive set-up (high-speed cameras, strong illumination, etc) as well as time-consuming post-processing. 
Other common Lagrangian techniques, \emph{e.g.}~particle tracking velocimetry (PTV), generally do not allow a direct measurement of the possible rotation of the particle simultaneously with its translation.

The experimental technique presented here was designed to overcome these issues thanks to the design of \emph{instrumented particles}~\cite{Gasteuil:2007tw,Gasteuil:2009la,Shew:2007cy,Pinton:fk}.
This was initially developed by our group to study Lagrangian particles with a temperature sensitive dependance, there used in Rayleigh-B\'enard convection~\cite{Xia:2012}. 
The approach is to instrument a neutrally buoyant particle in such a way that it measures the temperature as it is entrained by the flow, and to  transmit the data via a radio frequency link to the lab operator. 
This way, one gains access to trajectories for as long as the particle's battery lifetime.
In the work reported here, we built upon this approach to instrument the particle with a 3D accelerometer such that one gets the accelerations -- \emph{i.e.} the forces -- acting on a spherical particle  in real time and for long trajectories.
The instrumented particle has been previously tested, benchmarked and validated with the optical technique of Zimmermann \emph{et al.}~\cite{Zimmermann:2012fk}, showing a good agreement between the two different measurements of the acceleration.
In the present work we establish methods to extract physical  characteristics of the investigated flows  from the particle's acceleration signal.

One further motivation is to gain insights into a flow when direct imaging is not possible, \emph{e.g.}~when dealing  with opaque vessels, non-transparent fluids or granular media. These constraints occur especially in industry where additional bio-medical or environmental constraints arise (the injection of tracer particles might  be  unsuitable and thus prevent any visualization technique). 
As mentioned above solid particles are found to rotate  when advected in a highly turbulent flow~\cite{liftforce}.
We show here that it is possible to build quantities depending or not on the particle's rotation and we conclude about  flow parameters that are directly accessible without any optical measurement.

The article is organized as follows:
first, we present the experimental setup, as well as a brief reminder of the technical characteristics of the instrumented particle and the forces it measures  (section~\ref{sec:setup}).
Then, we present the new signal processing methods (section~\ref{sec:signals}).
Finally, we discuss and conclude on this new measurement technique (section~\ref{sec:discussion}).

\section{Experimental setup}
\label{sec:setup}

\subsection{Instrumented particle}

\begin{figure*}[tb] 
\centering
      \includegraphics[width=0.8\textwidth]{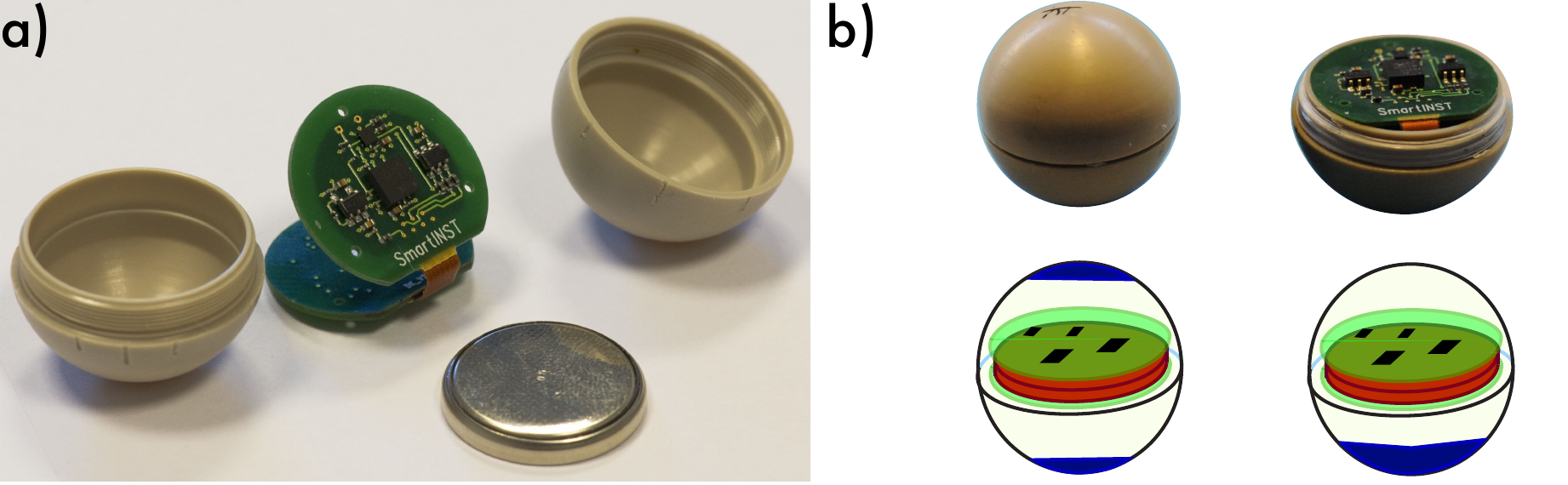} 
   \caption{a)~Picture of the instrumented particle (so-called smartPART from smartINST S.A.S.); \quad b)~Possible mass distributions of the particle; its inertia consists mainly of a disk and a spherical shell, with different density adjustment and imbalance settings by adding Tungsten paste (in blue); experiments are best done with a symmetrical repartition of the masses.}
   \label{fig:smartPartCircuit}
\end{figure*}

The device described in the following is designed and built by smartINST S.A.S., a spin-off from CNRS and the ENS de Lyon. 
It consists of an instrumented particle (the so-called smartPART \textregistered), a spherical particle which embarks an autonomous  circuit with 3D-acceleration sensor, a coin cell and a wireless transmission system, and a data acquisition center (the so-called smartCENTER \textregistered) which receives, decodes, processes and stores the signals from the smartPART (see \fig{fig:smartPartCircuit} and \fig{fig:vankarman}).
The smartPART measures the three dimensional acceleration vector $\aSP$ acting on the particle in the flow. It is in  good agreement with other techniques, details can be found in Ref.~\cite{Zimmermann:2012fk}.
The accelerometer consists of a micro-electro-mechanical system giving the three components of the acceleration (each of the three decoupled axes  returns  a voltage  proportional to the force acting on a small mass-load suspended by micro-fabricated springs). From this construction arises a permanent measurement of the gravitational force/acceleration $\vec g\equiv 9.8 \, \meter/\second^2 \cdot\vec {\hat e}_z= g\cdot  \vec {\hat e}_z$. 
Each axis has a typical full-scale range of $\pm 3.6 \,g=35 \, \meter/\second^2$. %data points exceeding this values are excluded from the analysis 
The sensor has to be calibrated in order to compute the physical accelerations from the voltages of the accelerometer. The detailed procedure is described in Ref.~\cite{Zimmermann:2012fk}. Concerning the resolution of the smartPART, the  uncertainty on the acceleration norm is  $\abs{\vec\sigma}=\sqrt{\sum_i\sigma_i^2}=0.008\,g$, with an average noise $\sigma_i\leq0.005\,g$ on each axis.

The particle rotates freely and in an  \emph{a priori} unknown way as it is advected by the flow.
The instantaneous orientation of the particle can be described by an absolute orientation with respect to a reference coordinate system, $\matrize R(t)$~\cite{goldstein}. 
For readability we only write the time reference when necessary (\emph{e.g.}~when two different times are involved in an equation) and drop it otherwise.
Using this rotation matrix, it is possible to express the contributions to the force acting on the particle and measured by the acceleration sensor in the lab frame or in the particle frame.
The following contributions to the particle's acceleration signal $\aSP$ can be identified:
\begin{description}
\item[(i) Gravity:] By construction, gravity $\vec g$ is  always contributing to $\aSP$. Since the particle is \emph{a~priori} oriented arbitrarily in space, $\vec g$ is projected onto all 3 axes.
\item[(ii) Translation:] The forces acting on the particle are projected as the Lagrangian acceleration $\displaystyle{\vec \atrans= \frac{\d^2}{\d t^2}\vec x(t)}$  onto the sensor. %However the projection changes if the sensor is rotating. %\displaystyle{\frac{d^2}{dt^2}\vec x(t)}
\item[(iii) Rotation:] The particle rotates freely around its geometrical center with an angular velocity $\vec \omega$. If the sensor is placed by $\vec r$ outside the geometrical center of the sphere one observes the centrifugal force: $\displaystyle{\vec a_\text{cf}=\vec \omega \times \kla{\vec \omega \times \vec r} + \left(\frac{\d}{\d t}\vec \omega\right) \times \vec r}$. According to the technical drawings it is $\vec r\approx 3\mm\cdot \hat e_z$. 
Experiments on the rotation of the smartPART in a \karman flow created by two counter-rotating impellers show that the angular velocity $\omega$ of the particle is of the order of the impeller frequency \fimp~\cite{liftforce,Zimmermann:2012fk}. The rotational forces  are of order $\displaystyle{r \,\omega^2\sim{r}\cdot({4\pi^2\,\fimp^2})  \lesssim0.1\,\abs{\atrans} }$  and have consequently negligible effect.
A more detailed analysis measured that  the ratio between the contribution due to the rotation and the total acceleration  is $\displaystyle{{\abs{\vec a_\text{cf}}}/{\abs{\atrans+\vec a_\text{cf}}}<0.1}$~\cite{Zimmermann:2012fk}.
The contribution due to the rotation is thus neglected.
It has to be noted that by construction of the accelerometer and because the circuit is fixed within the sphere, there is no contribution of the Coriolis force.
\item[(iv) Noise  \& spikes:] In ideal situations the smartPART has a noise of less than $0.005\,g$ for each axis, which can be handled by a moving average. Wrong detections appear as strong deviations from the signal and are hard to distinguish from high acceleration events due to the turbulent flow or contacts with \emph{e.g.}~the impellers. Experiments in different configurations prove the remaining noise to be negligible~\cite{Zimmermann:2012fk}.\\
\end{description}
Combining the different terms, and neglecting possible noise and the rotational bias yields:
\begin{equation}\begin{split}
 \aSP &\approx\,\matrize R \,\bra{ \vec g + \frac{\d^2\;}{\d t^2}\vec x} =  \matrize R \, \left[ \vec g + \atrans \right] .
\end{split}\label{eq:accJim}\end{equation}
The contributions due to  gravity and  translation are thus entangled by the continuously changing orientation of the particle.
Since gravity is of little interest here, one has to investigate how common quantities such as the mean and the variance (or \abc{rms}) of the acceleration time series as well as  auto correlation functions can give information about the particle motion.\\

Considering  robustness, the smartPART is able to continuously transmit data for a few days. During various experiments in a \karman flow neither contacts with the wall nor with the sharp edged blades of the fast rotating impellers damaged its function or shell. 
Furthermore, the sensor has among other things been chosen for its weak temperature dependance; in order to achieve optimal  precision of the measurements, we calibrate the particle at experiment temperature shortly before the actual experiment.
%Owing to the continuous data transmission of the instrumented particle, one flow configuration can be characterized in approximately 30 minutes.

Finally, by adding Tungsten paste to the inside of the smartPART the weight of the particle can be adjusted such that the particle is neutrally buoyant in de-ionized water at $20\,\degreecelsius$.
It  should be noted that the mass distribution inside the particle is neither homogeneous nor isotropic. 
The particle's  inertia is best described by a heavy disk of $20\mm$ diameter (the battery), a spherical shell and  patches of Tungsten paste. 
The paste must, therefore, be added carefully to minimize the imbalance of the particle (see Fig.~\ref{fig:smartPartCircuit}b); otherwise the resulting out-of-balance particle (\emph{i.e.}~with the center of mass not coinciding with the geometrical center) induces a strong preferential orientation and wobbles similar to a kicked physical pendulum. 
For a well balanced particle, which  rotates easily in the flow, one of the eigen-axes  of inertia then coincides (approximately) with the  $z-$axis of the accelerometer. The other two are within the $x-y$ plane due to  rotational symmetry.

\subsection{\Karman swirling flow}

\begin{figure*}[tb] %  figure placement: here, top, bottom, or page
   \centering
   \includegraphics[width=0.99\textwidth]{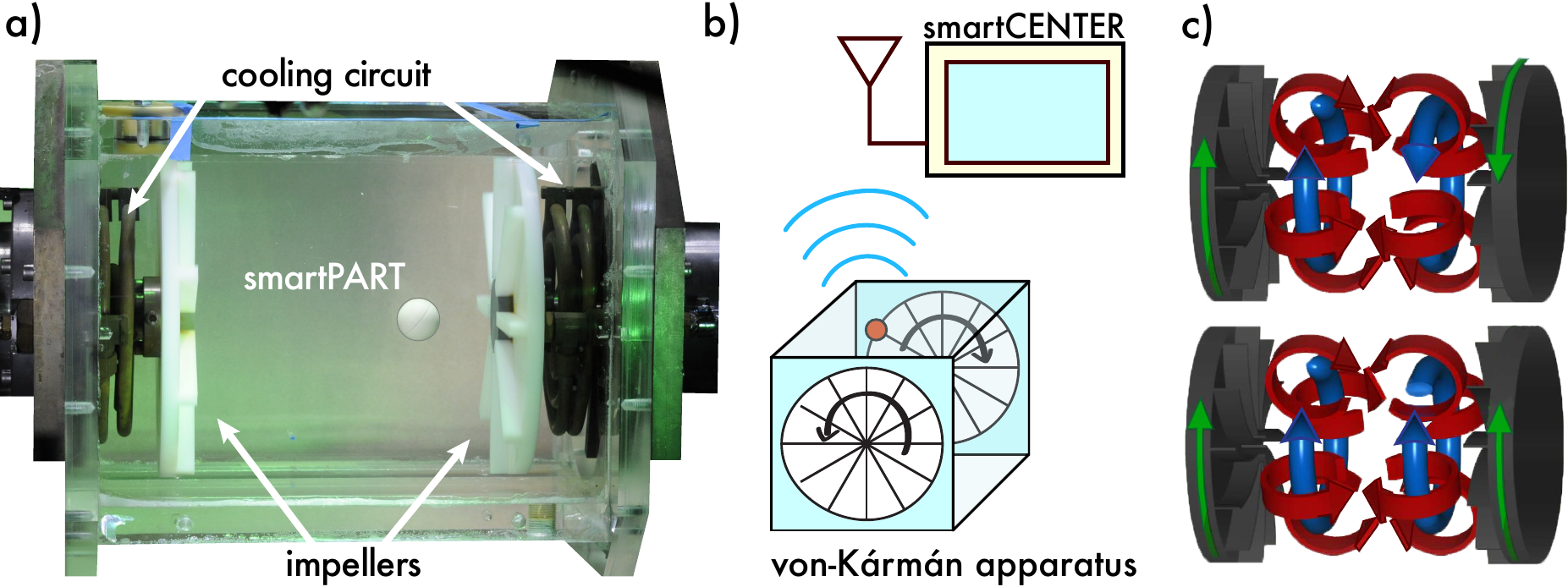} 
   \caption{a)~Picture of the \karman swirling flow with the instrumented particle inside; \quad b)~Sketch of the experimental setup with the apparatus and the smartPART transmitting acceleration signals to the smartCENTER; \quad c)~Sketch of the global structures that can be found with the two co- and counter-rotating regimes.}
   \label{fig:vankarman}
\end{figure*}

We investigate the motion of the instrumented particle in a fully turbulent flow.
Namely, we use a \karman swirling flow; in contrast to  Ref.~\cite{liftforce} the apparatus is here filled with water and develops  higher turbulence rates.
A swirling flow is created in a square tank by two opposing counter-rotating impellers of radius $R=9.5\,\centi\meter$ fitted with straight blades $1\,\centi\meter$ in height  (see Fig.~\ref{fig:vankarman}). 
The flow domain in between the impellers has characteristic length $H  = 20\,\centi\meter \cong 2R$.
Blades on the impellers work similar to a centrifugal pump and add a poloidal circulation at each impeller.  
For counter-rotating impellers, this type of flow is known to exhibit fully developed turbulence~\cite{RAVELET:2008ja}.
Within a small region in the center  the mean flow is little and  the local characteristics approximate homogeneous turbulence. 
However, at   large scales  it is known to be anisotropic~\cite{Ouellette:2006fk,Monchaux:2006fj}. 
Key parameters of the turbulence at different impeller speeds are given in Table~\ref{tab:characteristics}.
The two impellers can also be driven co-rotating, creating a highly-turbulent flow inside the vessel with one pronounced persistent global vortex along the axis of rotation. 
Close to the axis of rotation the mean flow is weak, followed by a strong toroidal component and an additional poloidal circulation induced by blades on the impellers.  The energy injection rate is a factor of 2 smaller than for counter-rotating impellers. This means that at the same impeller frequency, \fimp,  co-rotating driving creates less turbulence than  counter-rotation, but the flow is still highly turbulent~\cite{these:catherine}. 
Although the vortical structures near the disks are comparable (see Fig.~\ref{fig:vankarman}c), the co- or counter-rotating regimes yield well distinct global structures in the center of the vessel. The two regimes are used to compare the signals obtained by the instrumented particle in two very different flow configurations. In addition, the co-rotating serves as  a test case for persistent, large vortex structures as they are found in mixers with only one impeller.
%\NOTE{hervorheben dass corot == largescale vortex}

\begin{table}[t]
   \centering
   \small
   \begin{ruledtabular}
   \begin{tabular}{ r r r r r r r} 
    \multicolumn{1}{r}{$\fimp [\hertz]$}  & \multicolumn{1}{c}{$Re$}& \multicolumn{1}{c}{$R_\lambda$}& \multicolumn{1}{c}{$\eps [\meter\squared\per\second\cubed]$} & \multicolumn{1}{c}{$\eta[\micro\meter]$}& \multicolumn{1}{c}{$\tau_\eta [\milli\second]$}& \multicolumn{1}{c}{$T_\text{int} [\second]$}  \\\toprule
  $1.0$ & $62800$ & $290$ & $0.07$ & $62$ & $3.8$ & $1.00$\\
  $2.0$ & $125700$ & $410$ & $0.48$ & $38$ & $1.4$ & $0.50$\\
  $3.0$ & $188500$ & $505$ & $1.68$ & $28$ & $0.8$ & $0.33$\\
  $4.0$ & $251300$ & $580$ & $4.03$ & $22$ & $0.5$ & $0.25$\\
      \end{tabular}
      \end{ruledtabular}
   \caption{\small Key-parameters of the counter-rotating flow configuration. The integral time scale is defined as $T_\text{int}=1/\fimp$ and the integral length scale is estimated to be $L_\text{int}=3\cm$. 
   We use the following definition for the Reynolds numbers:  $Re=2\pi R^2\fimp/\nu$ and $R_\lambda\approx\sqrt{\frac{15}{\nu}\cdot 2\pi L_\text{int}^2\fimp}$. Note that the  particle explores the whole apparatus, where the  flow is known to be inhomogeneous and anisotropic. Thus,  $R_\lambda$ and the Kolmogorov scales are only  rough estimates. 
   For comparison: co-rotating impellers yield an energy injection rate which is half of the energy injection rate of counter-rotating impellers at the same impeller frequency.}
   \label{tab:characteristics}
\end{table}

%& \multicolumn{1}{c}{$\fprop [\hertz]$}& \multicolumn{1}{c}{$\Rey$}& \multicolumn{1}{c}{$R_\lambda$}& \multicolumn{1}{c}{$\eps [\meter\squared\per\second\cubed]$} & \multicolumn{1}{c}{$\eta[\micro\meter]$}& \multicolumn{1}{c}{$\tau_\eta [\milli\second]$}& \multicolumn{1}{c}{$\Tint [\second]$}\\\toprule
%\multirow{6}{*}{\begin{sideways}counter-rot\end{sideways}} & $0.5$ & $31400$ & $205$ & $-0.01$ & $71$ & $0.0$ & $2.00$\\
% & $1.0$ & $62800$ & $290$ & $0.07$ & $62$ & $3.8$ & $1.00$\\
% & $2.0$ & $125700$ & $410$ & $0.48$ & $38$ & $1.4$ & $0.50$\\
% & $3.0$ & $188500$ & $505$ & $1.68$ & $28$ & $0.8$ & $0.33$\\
% & $4.0$ & $251300$ & $580$ & $4.03$ & $22$ & $0.5$ & $0.25$\\

%%%%%%%%%%%%%%%%%%%%%%%%

\section{Acceleration signals}
\label{sec:signals}
\begin{figure}[tbh] %  figure 
   \centering
   \includegraphics[width=0.9\textwidth]{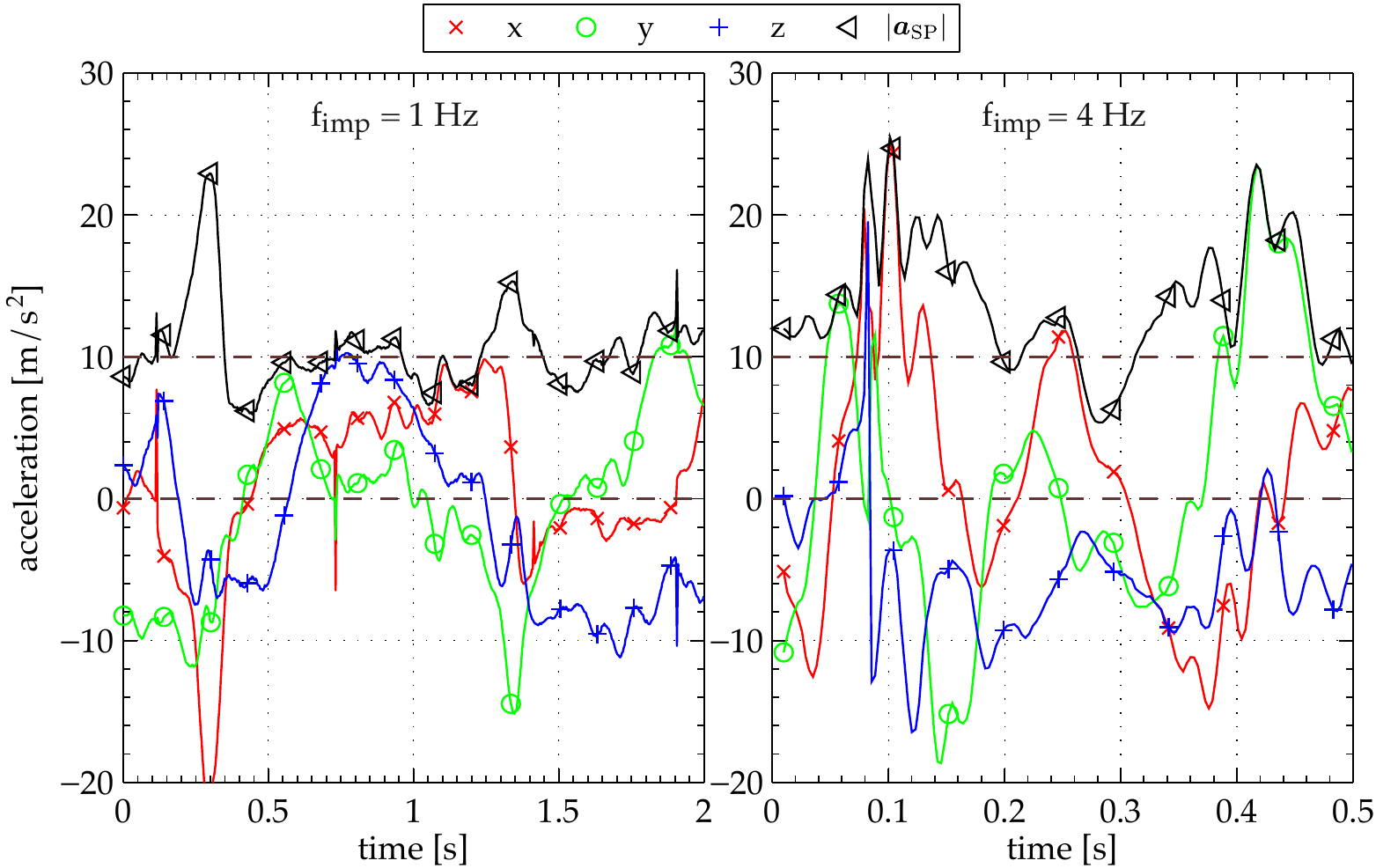} 
   \caption{Sample acceleration time-series, $\aSP(t)$, with the impellers counter-rotating at $\fimp=1\Hz$ (left) and $\fimp=4\Hz$ (right). Both samples last 2 integral times $T_\text{int}$.}
   \label{fig:trace}
\end{figure}

\fig{fig:trace} shows two sample time-series of the three components of the acceleration measured by the instrumented particle in the \karman flow, superimposed with the norm of the acceleration.
Two different frequencies of the impellers are presented here: $1\,\hertz$ and $4\,\hertz$.
The mean value of the norm fluctuates around $g=9.8 \,\meter\per\second\squared$, indicating that gravity is always measured by the accelerometer.
Furthermore, the fluctuations of the norm increase with the impeller frequency.
It is, however, difficult to compare the three components of the acceleration, either between each other or for different impeller frequencies.
This is mainly due to the measurement of the gravity that is randomly projected on the three axes of the accelerometer as the particle rotates in the flow.
It results in signals containing both contributions from the gravity and the particle's translation, with no straightforward method to separate them.
In other words, contrary to other methods (\emph{e.g.}~particle tracking velocimetry) it is not possible to obtain directly the characteristics of the particle motion.
Hence, one needs to post-process  the data to derive information about the statistics and the dynamics of the particle.\\

\subsection{Analysis of the raw signal $a_\text{SP}$}

\begin{figure*}[tbh] %  figure placement: here, top, bottom, or page
   \centering
   \includegraphics[width=0.99\textwidth]{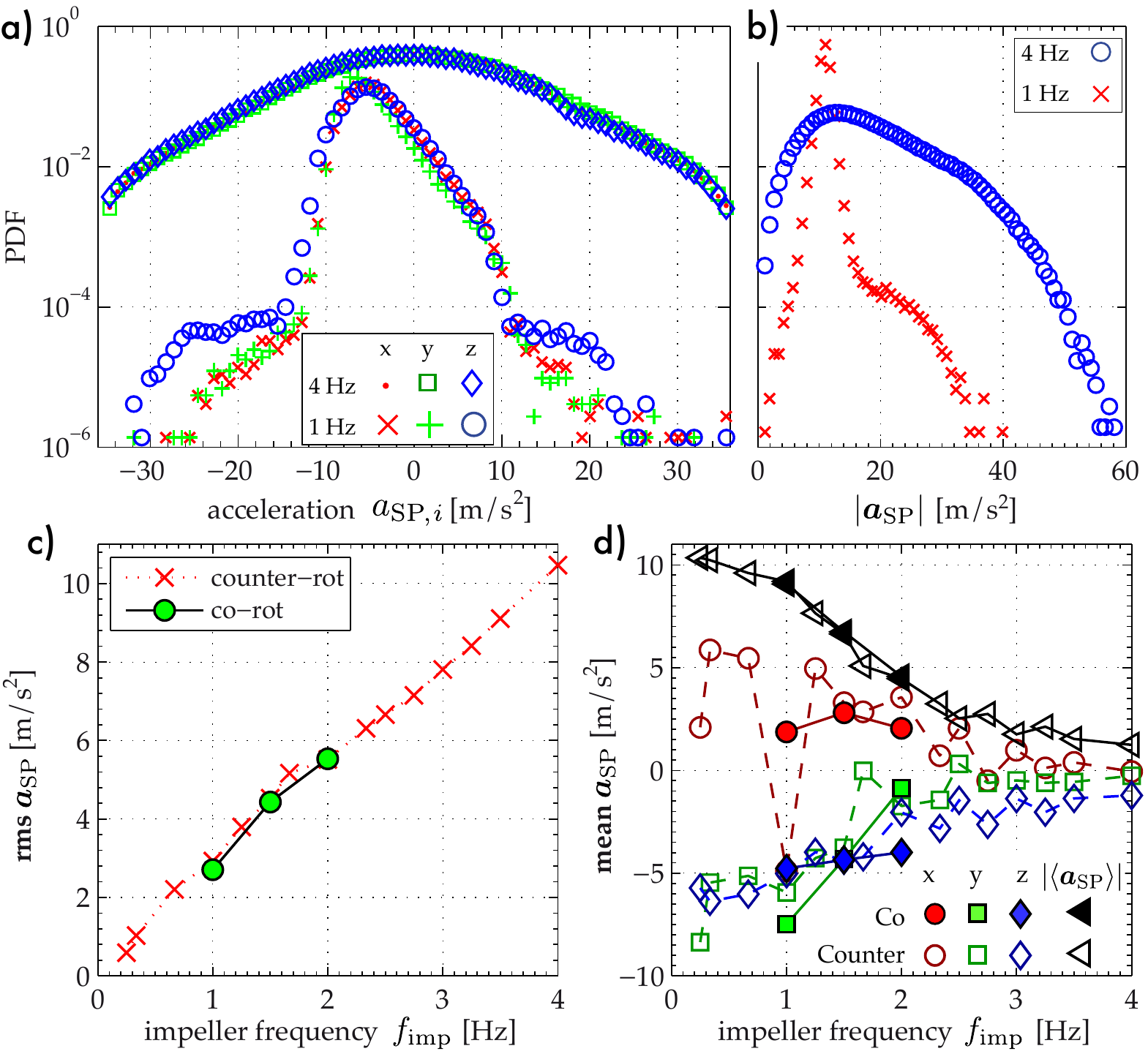} 
   \caption{Top:~Probability density functions of a)~the components and b)~the norm of the acceleration $\aSP$ for two different impeller frequencies. For readability, the PDFs of acceleration components at $4\,\hertz$ have been arbitrarily shifted. \quad Bottom:~Evolution of c)~$\abc{rms}\,{\aSP}$ and d)~$\mean{\aSP}$ with the impeller frequency $\fimp$; filled symbols  ($\circ$) indicate co-rotating impellers. In all cases the particle explored the flow for a sufficient amount of time for the statistics to converge.
    In good agreement with \gleich{eq:aSPmean}, $\abs{\mean{\aSP}}$  continuously decreases  from $1 g$ to $0 g$ as the impeller frequency increases.}
   \label{fig:meanAsp}
\end{figure*}

\fig{fig:meanAsp} shows different results of a basic statistical analysis of the acceleration signals, namely the PDFs of the three components and the norm of the acceleration for different impeller frequencies, and the fluctuating and mean values of the acceleration as a function of the impeller frequency.
The accelerometer used in the smartPART saturates if one of the acceleration components exceeds $\pm 3.6g$, we exclude these points from the analysis. 
This removal diminishes the observed acceleration and the bias increases with the forcing. 
In the case of \fig{fig:meanAsp}, almost $3\%$ of all data points were removed at $\fimp=4\Hz$, which is two orders of magnitude higher than for $\fimp=1\Hz$. 
Looking at the PDFs of the acceleration for a given impeller frequency (\fig{fig:meanAsp}a), one can see that the three components give similar results for a wide range of acceleration values.
However, the PDFs are very different from one frequency to another.
Whereas at low impeller frequencies the PDFs are skewed and shifted, they become centered and symmetric with increasing impeller frequency. 
This evolution in shape can be explained by the particle's mass distribution and imbalance.
Although the particle is carefully prepared, its moment of inertia is not that of a solid sphere and the particle's center of mass does not perfectly coincide with its geometrical center. Consequently, the  particle becomes slightly out-of-balance, with a preferred orientation at low impeller frequency: the peaks then correspond to the projection of $\vec{g}$ on the axes in this preferred orientation and fluctuations around it.
When the impeller frequency (and consequently the turbulence level) increases, the particle is able to explore all the possible orientations, meaning $\vec{g}$ is randomly projected in all directions, and the asymmetry disappears.
The PDFs of the norm $\abs{\aSP}$ (\fig{fig:meanAsp}b) also show this difference in shape, with a clear peak near the value $g$ (again, gravity is always measured by the 3D accelerometer), but with a narrow strong peak at low impeller speed and a more stretched PDF at high impeller speed.

The evolution of the fluctuations of acceleration (\abc{rms} \aSP) with the impeller frequency is given in \fig{fig:meanAsp}c.
Only one component of the acceleration is presented here for readability, since no preferred direction in any of the axes was found.
This results in all three components \abc{rms} values having the same behavior and amplitudes.
One can see that surprisingly, the fluctuations of acceleration increase linearly with the frequency.
That is in contrast to dimensional arguments that tell $\atrans  \propto \fimp^2$. 
Moreover, it is not possible to distinguish between the co- and counter-rotating regimes of the impellers.

\fig{fig:meanAsp}d shows $\mean{ \aSP }$ as a function of the impeller frequency, $\fimp$, and the forcing. 
As expected the mean accelerations are  becoming smaller with increasing impeller frequency. 
Indeed taking the average of  \gleich{eq:accJim} yields:
\begin{equation}\begin{split}
\mean{ \aSP } = \mean{ \matrize R \, \vec g } + \mean{ \matrize R \, \vec \atrans  }.
\end{split}\label{eq:aSPmean}\end{equation}
If the particle  explores continuously all the possible orientations, the mean vanishes; whereas a fixed orientation (\emph{i.e.}~no rotation) yields $\mean{ \aSP } =  \matrize R \left(\vec g +  \mean{  \vec \atrans  }\right)$.
This is what is observed in \fig{fig:meanAsp}. 
In the case of weak turbulence (\emph{i.e.}~for smaller values of $\fimp$), the mean acceleration gives an estimate of gravity: $\mean{\aSP}\approx\mean{\matrize R \, \vec g}$.
However, for stronger turbulence and even if the mass distribution slightly induces a preferred direction, the particle can rotate freely around this axis, resulting in a vanishing mean acceleration when the impeller frequency increases: $\mean{\aSP}\rightarrow 0$.
In the latter case, contacts with impellers, walls, eddies, etc also help overpowering any preferred direction easily and force the particle to rotate. 
It can be noted that again, it is not possible to distinguish between the co- and counter-rotating regimes.
Furthermore, the variance of a component $a_\text{SP,i}$ of $\aSP$ depends strongly on its mean value, $\mean{a_\text{SP,i}}$. 
As explained before, gravity renders $\mean{a_\text{SP,i}}$  non-negligible. 
Additionally, we observe for weak turbulence levels ($\fimp\lesssim1\Hz$)  that particles  are able to stay in an orientation for several seconds. 
Hence, a global mean of the complete time-series is not a meaningful quantity. 

The direct study of the raw acceleration signal,   $\aSP$, only allows to conclude whether the particle rotates or not.
It does not permit to disentangle the contributions of the gravity and the particle translation, and subsequently to have any precise insight on the flow.
Other methods adapted to this problem are thus needed to extract informations from the instrumented particle related to  its motion.

\subsection{Moments of the acceleration due to the particle's translation} % for some reason \vec command crashes the compilation in \subsection
\label{ss:SPmoments}
In confined flows and provided the statistics are converged, it is $\mean{\atrans }=  0$. 
One is, therefore, interested in the PDF of $\atrans $. 
Although, we mentioned we don't have direct access to $\atrans $ and its PDF, we can compute the even (central) moments of its PDF.\\

\begin{figure}[tb] %  figure placement: here, top, bottom, or page
   \centering
   \includegraphics[width=0.8\textwidth]{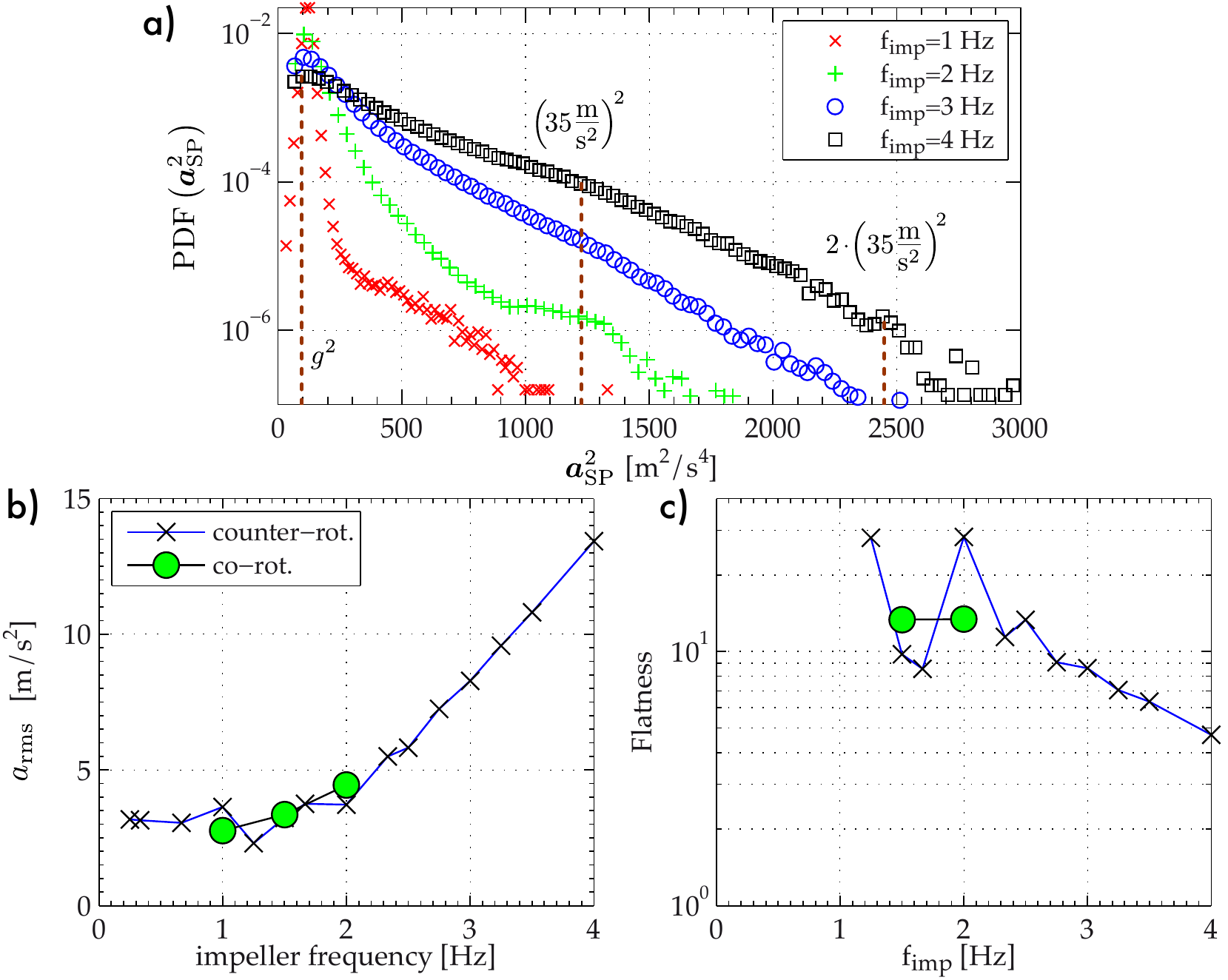} 
   \caption{Moments of $\atrans$. a) $\PDF\kla{\aSP^2}$ at different impeller frequencies. The 3 vertical lines mark gravity and the saturation of one or two accelerometer axes. b) and c) : RMS  and fourth  moment of $\atrans$ as a function of the impeller frequency. $\acct(\fimp)=\sqrt{\mean{\abs{\aSP(\fimp)}^2}-g^2}$.}
   \label{fig:Atrans}
\end{figure}

The variance of $\aSP$ is 
\begin{equation}\begin{split}
 \mean{\aSP^2}&=\mean{\abs{\matrize R \, \vec g}^2} +\mean{\abs{\matrize R \,  \atrans }^2 } + 2\mean{\matrize R \, \vec g \cdot \matrize R \,  \atrans } \\
&=g^2 + \mean{\atrans^2} + 2\,g \mean{ a_z},
\end{split}\label{eq:Amagn1}\end{equation}
where $a_z\equiv \vec {\hat e}_z \cdot \vec \atrans $.
It should be kept in mind, that each axis of the smartPART's accelerometer is limited to $\pm3.6\,g$, and possible events of higher acceleration are therefore not included in the analysis.
The PDF of $\abs{\aSP}^2$ for different impeller frequencies is shown in \fig{fig:Atrans}a.
As expected, a peak is clearly observed at $g^2$.
One can also see that there are breaks in the slope at $\abs{\aSP}^2\approx(3.6\,g)^2$  and $\abs{\aSP}^2\approx2\,(3.6\,g)^2$, corresponding to cases where one or two axes would saturate. 
Some information is inevitably lost, and to investigate the  behavior at large $\fimp$, the sensor would have to be  replaced with a different model supporting higher accelerations.\\

If the particle is neutrally buoyant and the flow is confined, one expects $\mean{a_z}= 0$.
We therefore obtain an estimate of the standard deviation (\abc{rms}) of $\atrans $:
 \begin{equation}\begin{split}
\acct\equiv \sqrt{ \mean{\atrans^2} }=\sqrt{\mean{\aSP^2}-g^2}.
\end{split}\end{equation}
$\acct$ is independent of how gravity is projected on the axes of the accelerometer (in other words it is insensitive to the particle's absolute orientation).
Nevertheless,  a bad calibration (\emph{e.g.}~caused by longterm drift or strong temperature change) can  introduce a systematic offset to $\acct$. Nevertheless, this bias can be minimized by calibrating the thermalized smartPART before the actual experiment. 
\fig{fig:Atrans}b depicts the evolution of $\acct$ with the impeller frequency. 
In agreement with dimensional analysis, $\acct\kla{\fimp}$ describes a parabola, although one could also argue that  $\acct$ seems linear with ${\fimp}$ for $\fimp\geq3 \,\hertz$. As illustrated in Fig.~\ref{fig:Atrans}a, this is caused by  saturation of the accelerometer, which cuts off/underestimates high acceleration events present at these high turbulence level.\\

Similarly to the variance, one can estimate the fourth central moment of $\atrans $. It is
\begin{equation}\begin{split}
 \meanB{ \abs{\aSP}^4} \;&=\;  \mean{ \bra{g^2 +\atrans^2 +2\, g\,a_z } \bra{g^2 +\atrans^2 +2 \, g\,a_z }} \\%[3mm]
\quad&= g^4+ \meanB{\abs{\atrans}^4} + 2 \,g^2\, \mean{\atrans^2}+  4\, g^2 \mean{a_z^2}
+4\,g^3\mean{ a_z} + 4 \, g\mean{\abs{\atrans }^2   a_z}.
\end{split}\label{eq:flatness1}\end{equation} 
Assuming no preferred direction in $\vec \atrans$, as found for small particles in a windtunnel~\cite{Voth:2002hc} and verified for solid particles of size comparable to the integral length scale   in the same apparatus~\cite{Zimmermann:2012th}, one has   $4 g^2 \mean{a_z^2}\approx 4/3 \, g^2\,  \atrans ^2$. 
Again, the terms $\mean{a_z}$, $4\,g^3\mean{ a_z}$ and $4 g \meanB{\abs{\vec \atrans }^2   a_z}$ are expected to vanish in the case of confined flows. 
\gleich{eq:flatness1} then simplifies to
\begin{equation}
\meanB{ \abs{\aSP}^4} \approx g^4 + \meanB{\abs{\atrans}^4}+ \frac{10}{3} g^2 \acct^2. 
\label{eq:flatness2}\end{equation} 
The flatness, $F(\atrans )$, is defined as: 
\begin{equation}\begin{split}
F\kla{\atrans}=\frac{\meanB{\abs{\atrans}^4}}{\mean{\atrans^2}^2}=\frac{\meanB{ \abs{\aSP}^4} - g^4- \frac{10}{3} g^2 \acct^2  }{\acct^4}.
\end{split}\end{equation}
As shown in \fig{fig:Atrans}c we observe a flatness of the order of $10$ in our \karman flow, which is close to the flatness obtained in the case of much smaller particles~\cite{qureshi2007turbulent} and to our finding for solid particles of similar size~\cite{Zimmermann:2012th}.
The uncertainty in the flatness can partially be attributed to an uncertainty in $g$ and stems from the resolution, noise and measurement range of the smartPART, but also from the particle's weak drift. It is furthermore biased by contacts with  impellers and walls. More surprisingly, the flatness  decreases with the forcing.  
This decline is again due to the limited measurement range of the accelerometer used: at high accelerations the sensor saturates and  thereby sets $\PDF\klaB{\atrans  \,\big|\, \abs{a_{\text{SP,}i}}>3.6g}=0$. 
Since the flatness is the fourth moment of the PDF and as such highly sensitive to strong  accelerations, we find a decrease whereas solid large spheres in the same flow have an increasing flatness~\cite{Zimmermann:2012th}.
We also conclude that calculating moments of higher order is out of reach.  \\
%Albeit the force is a three dimensional quantity we derive scalar information...weird isn't it

It is remarkable, that based only on the second and fourth  moment  of \atrans~  one cannot  clearly distinguish between a counter-rotating and a co-rotating flow although these two forcings induce two clearly different large scale flow structures. Similar behavior has been found for solid spheres of comparable size in the same flow~\cite{Zimmermann:2012th}, too.
It should be noted that  the energy injection rates for the two ways of driving the flow  differ  by only a factor of 2: the co-rotating forcing is highly turbulent, too. In addition, in vicinity of the disks the flow has a strong contribution of the centrifugal pumping of the  blades on the impellers and the flow configurations are comparable in that region.

%Although the flow in vicinity of the impellers  is comparable
% 
%It has to be noted that the two flows are created in a very similar way, with blades on rotating impellers that set the fluid in motion. The global structures are thus comparable near the impellers, with a change occurring near the center of the vessel: two counter-rotating global structures will create homogeneous turbulence in the center whereas two co-rotating structures will create a global structure at the scale of the vessel.
%If the particle stays close to the impellers, its dynamics will be mainly driven by the global structures present at this location. \NOTE{Muck}
%This renders the distinction between the two regimes impossible.

\subsection{Auto-correlation functions}
In order to distinguish between the two regimes, we now turn to the auto-correlation of the acceleration time-series to estimate  correlation time scales of the flow. 
Ideally, one would want to compute the auto-correlation of the translational force, \emph{e.g.} $\mean{\atrans \kla{t}\cdot \atrans \kla{t+\tau}}$, but again the constantly changing orientation of the smartPART blocks any direct access to $\atrans \kla{t}$ and quantities derived thereof. 
We therefore need to find quantities, which are either not altered by the orientation of the smartPART or extract information on  its rotation.

\subsubsection{An Auto-correlation  invariant to the rotation of the sensor}\label{sss:invariantAC}
\begin{figure*}[htb] %  figure AC trans
   \centering
   \includegraphics[width=0.8\textwidth]{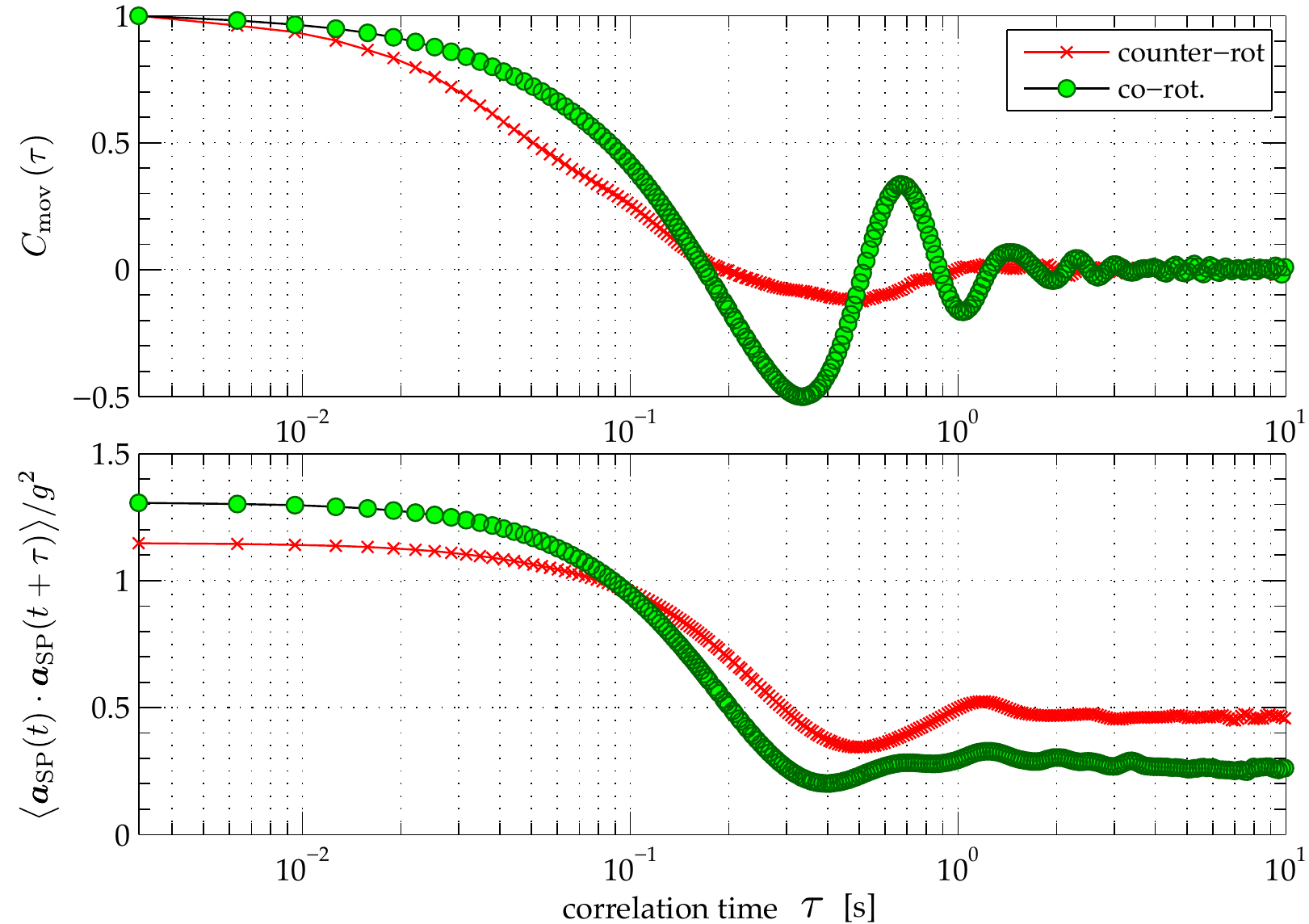} 
   \caption{Top:~Rotation-invariant auto-correlation function $C_\text{mov}\kla{\tau}$ (\gleich{eq:ACjim1} after rescaling); the auto-correlation significantly differs between counter- and co-rotating impellers.\quad Bottom:~Rotation-sensitive auto-correlation function $\mean{ \aSP(t)  \cdot \aSP(t+\tau)}/g^2$. In all cases the impeller frequency is $1.5\Hz$. A logarithmic scale  is chosen for the abscissae to display both short and long time contributions to the correlations. }
   \label{fig:acMagn}
\end{figure*}
In the spirit of \gleich{eq:Amagn1} and \gleich{eq:flatness1} one can construct the auto-correlation function of the magnitude of $\aSP$. It is:
\begin{equation}\begin{split}
& \mean{ \abs{\aSP(t)}^2   \abs{\aSP(t+\tau)}^2} \\%[3mm]
 &\quad=\;  \mean{ \bra{g^2 +\atrans ^2(t) +2\, \vec g \cdot  \atrans (t)} \bra{g^2 +\atrans ^2(t+\tau) +2 \,\vec g \cdot  \atrans (t+\tau)}} \\%[3mm]
% &\quad= \mean{\abs{ \atrans (t)}^2 \abs{ \atrans (t+\tau)}^2} + g^4 + 
% g^2\kla{ \mean{\atrans^2 (t)}+ \mean{\atrans^2 (t+\tau)}}\\
% &\qquad  +2\,g^2\klaB{\mean{\vec g \cdot  \atrans (t+\tau)}+ \mean{\vec g \cdot  \atrans (t)}} + 2 \mean{\abs{ \atrans (t)}^2 \vec g \cdot  \atrans (t+\tau)}\\
%&\qquad + 2\mean{\abs{ \atrans (t)}^2 \vec g \cdot  \atrans (t+\tau)}  +  4 \mean{\kla{\vec g \cdot  \atrans (t+\tau)}\kla{\vec g \cdot  \atrans (t)}}\\%[3mm]
&\quad=\mean{\abs{ \atrans (t)}^2 \abs{ \atrans (t+\tau)}^2} + g^4 + 2 g^2\, \mean{\atrans^2}+  4 g^2 \mean{{  a_z(t)}\,{a_z(t+\tau)}}\\
 &\qquad  +4\,g^3\mean{ a_z} + 2 g \mean{\abs{ \atrans (t+\tau)}^2   a_z(t)} + 2 g \mean{\abs{\atrans (t)}^2   a_z(t+\tau)}  \\%[3mm]
  &\quad \approx \mean{\abs{\atrans (t)}^2 \abs{ \atrans (t+\tau)}^2} +  g^4+2\,g^2 \acct^2  + 4 g^2 \mean{{  a_z(t)}\;{a_z(t+\tau)}} .
\end{split}\label{eq:ACjim1}\end{equation} 
Again, the terms containing $a_z\equiv\vec {\hat e}_z \cdot \vec \atrans $ are expected to have zero mean. 
However, the last term on the right-hand side of \gleich{eq:ACjim1} does not vanish for $\tau\approx 0$, becoming $4 g^2 \mean{{  a_z(t)}\;{a_z(t+\tau)}} = 4 g^2 \mean{\abs{  a_z}^2}$. 
Assuming no preferred direction in $\atrans $ this can be approximated as $4/3 \, g^2\,  \acct^2$.
In contrast to \gleich{eq:ACjim1}, we preferably compute the autocorrelation of the fluctuations around the mean $\mean{\vec a^2_{SP}}$.
Hence, the autocorrelation of the norm can be negative. We further normalize the auto-correlation such that  it is  $1$ at $\tau=0$. After rescaling, we refer to this rotation-invariant quantity as $C_\text{mov}\kla{\tau}$.

\fig{fig:acMagn} displays $C_\text{mov}\kla{\tau}$ for  the co- and counter-rotating regime at an impeller frequency of $1.5\Hz$. %We identified that 
The auto-correlation of the counter-rotating forcing is well approximated by a sum  of  exponential decays. % or the transient function of a critical damped oscillator.
In contrast thereto, we  observe that co-rotating impellers correspond to an auto-correlation function showing a damped oscillation, \emph{i.e.} ~ the smartPART observes the longer  coherence in the large scale motion of the flow. 
This is in agreement with Eulerian measurements, where   pressure probes were mounted in a \karman flow: whereas the counter-rotating flow produces typical pressure spectra,  the same probe in the co-rotating case yields a spectrum which peaks at multiples of the impeller frequency. 
Similar behavior has been reported  for the magnetic field  in a  \karman flow~\cite{Volk:2006fk} filled with liquid Gallium (in that particular case the co-rotating regime was created by  rotating only one impeller).

Summing up, $\meanB{ \abs{\aSP(t)}^2   \abs{\aSP(t+\tau)}^2}$ is insensitive to the particular tumbling/rotational dynamics of the particle and it gives necessary information to  determine the type of flow. We also checked that this result is not altered by a possible imbalance of the particle.

\subsubsection{An Auto-correlation related to the tumbling of the particle}

One can further focus on the rotation of the particle by considering the product:
\begin{equation}
\setlength{\jot}{10pt} % affecting the line spacing in the environment
\begin{split}
& \mean{ \aSP(t)  \cdot \aSP(t+\tau)} \\ 
&\quad=  \mean{ \bra{ \matrize R\kla{t} \, \kla{ \vec g + \vec \atrans (t) }}  \cdot \bra{\matrize R\kla{ t+\tau} \, \kla{ \vec g + \vec \atrans (t+\tau) }} } \\
&\quad=  \mean{ \bra{  \matrize R^T\big(t+\tau\big)\cdot \matrize R\big(t\big) \, \left (\vec g + \vec \atrans (t) \right)}  \cdot \bra{\left (\vec g + \vec \atrans (t+\tau) \right) }}\\
&\quad=\mean{ \vec g \cdot \bra{ \matrize T(t,\tau)\vec g}  } + \mean{ \left[  \matrize T(t,\tau) \vec \atrans (t) \right]\cdot\vec \atrans (t+\tau)}  +   \mean{\bra{\matrize T(t,\tau)\vec g}\cdot \vec \atrans (t+\tau) }  +  \mean{\left[\matrize T(t,\tau)  \vec \atrans (t)\right ] \vec g}\\
&\quad\approx g^2\mean{ \hat  {\vec e}_z \cdot\bra{ \matrize T(t,\tau)\hat {\vec e}_z  }  } + \mean{ \bra{ \matrize T(t,\tau)\cdot \vec \atrans (t) } \vec \atrans (t+\tau)},
 \end{split}\label{eq:ACJim2}\end{equation} 
where the term 	$\matrize T(t,\tau)\equiv \matrize R^T\big(t+\tau\big) \matrize R\big(t\big)$ is a rotation matrix related to the instantaneous angular velocity $\vec \omega$ of the particle as explained in Ref.~\cite{Zimmermann:2011uu}. 
Again, the two terms containing products of $\vec g$ and $\vec a$ vanish if the particle is neutrally buoyant.
The term $g^2\mean{ \hat  {\vec e}_z \cdot\bra{ \matrize T(t,\tau)\hat {\vec e}_z  }  } $  is related to the tumbling of a spherical particle~\cite{Wilkinson:2011fk}.
In contrast to the other auto-correlation \gleich{eq:ACjim1}, one cannot subtract a mean value prior to computing  $\mean{ \aSP(t)  \cdot \aSP(t+\tau)}$. 
To estimate the ratio between $g^2\mean{ \hat  e_z \cdot\bra{ \matrize T(t,\tau)\hat e_z  }  } $ and $\mean{ \bra{ \matrize T(t,\tau) \vec \atrans (t) } \cdot \vec \atrans (t+\tau)}$ it is helpful to normalize by $g^2$. 

If $\mean{ \bra{ \matrize T(t,\tau)\cdot \vec \atrans (t) } \vec \atrans (t+\tau)}$ becomes uncorrelated, $\mean{ \aSP(t)  \cdot \aSP(t+\tau)}$ does not necessarily vanish. If uncorrelated ($\tau\gg T_\text{int}$):
\begin{equation}\begin{split}
  \mean{ \aSP(t)  \cdot \aSP(t+\tau)} &\cong g^2 \mean{\matrize R \kla{ t} \hat  {\vec e}_z \cdot \matrize R\kla{ t+\tau} \hat  {\vec e}_z} \\
  &= g^2 \mean{\matrize R\hat  {\vec e}_z}^2 \geqslant 0 
\end{split}\label{eq:plateau}\end{equation}
That means $\mean{ \aSP(t)  \cdot \aSP(t+\tau)}$ approaches a plateau whose height is determined by the average orientation of the particle. In analogy to $C_\text{mov}$, one can then subtract $g^2 \mean{\matrize R\hat  {\vec e}_z}^2$ and rescale $\mean{ \aSP(t)  \cdot \aSP(t+\tau)}$, which is termed $C_\text{tumb}\kla{\tau}$ in the following.

The lower plot in \fig{fig:acMagn} depicts $\mean{ \aSP(t)  \cdot \aSP(t+\tau)}/g^2$ for the two forcing regimes at an impeller frequency of $1.5\Hz$. 
%For comparison to $\mean{ \abs{\aSP(t)}^2   \abs{\aSP(t+\tau)}^2}$ only the autocorrelation function changed; the configuration and $\fimp$ are the same. 
In agreement with Eq.~\eqref{eq:plateau}, a plateau is reached for both forcings.
\begin{figure*}[tb] %  figure AC rot
   \centering
   \includegraphics[width=0.8\textwidth]{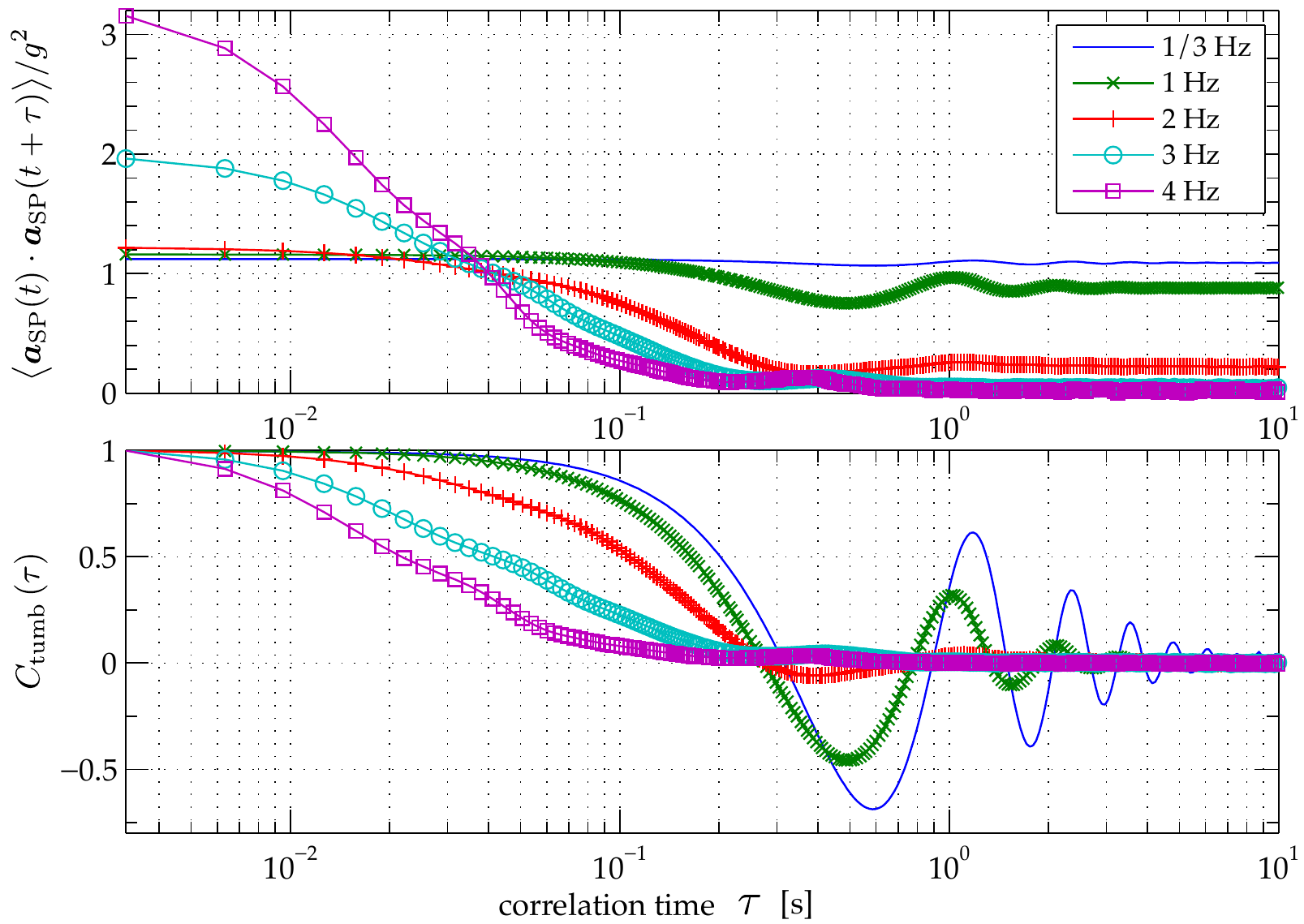} 
   \caption{Top:~Dependence of $\mean{ \aSP(t)  \cdot \aSP(t+\tau)}/g^2$ on the impeller frequency for counter-rotating impellers; \quad Bottom:~Same data after subtracting the plateau and rescaling.}
   \label{fig:acRot}
\end{figure*}
To investigate the role of the plateau we plot the auto-correlation of the particle for increasing $\fimp$ in \fig{fig:acRot}. 
For $\fimp\lesssim 1\Hz$ one finds little change with the plateau at almost $1$.
For $\fimp\approx 2\Hz$ the plateau drops but is still non-zero. The value of the plateau diminishes with further increase in \fimp. 
At the same frequency range we observe that the PDFs of the components of $\aSP$ become centered and symmetric (cf. \fig{fig:meanAsp}).
 $C_\text{tumb}\kla{\tau}$ -- \emph{i.e. } \gleich{eq:ACJim2} after subtracting the plateau and rescaling -- accesses the fluctuations around a mean value and is shown in the bottom plot of \fig{fig:acRot}. We find that $C_\text{tumb}\kla{\tau}$ evolves from a long-time correlated oscillatory shape at low impeller speeds to an exponential decay at high \fimp. %With increasing propeller speed $C_\text{tumb}\kla{\tau}$ 
%We have thus access to a second correlation  which is related to the rotation of the particle.
%Because \gleich{eq:ACJim2} depends on the rotation of the

\begin{figure*}[tb] %  figure placement: here, top, bottom, or page
   \centering
   \includegraphics[width=0.8\textwidth]{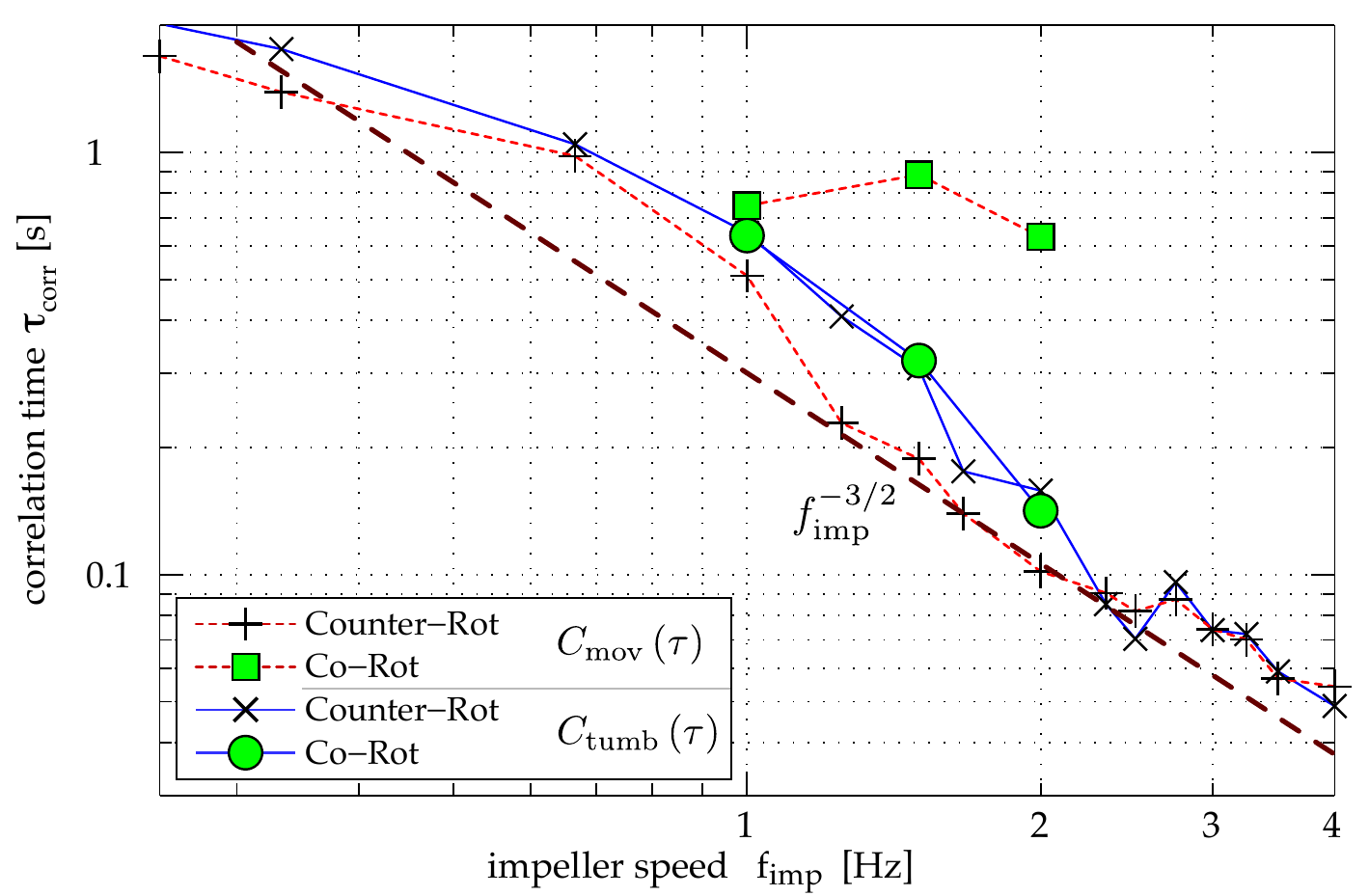} 
   \caption{Characteristic flow time scale \tcorr~determined from rotation-invariant, $C_\text{mov}\kla{\tau}$, and rotation-sensitive, $C_\text{tumb}\kla{\tau}$, auto-correlation function. Whereas both autocorrelation functions yield comparable \tcorr~ for counter-rotating driving, we find that their results are well distinct in the case of co-rotating impellers.  The dashed line indicates a $\fimp^{-3/2}$ power-law, as suggested by the scaling of the Kolmogorov time scale ($\tau_\eta\propto\eps^{-1/2}$ and $\eps \propto \fimp^3$).}
      \label{fig:timesJim}
\end{figure*}

\subsubsection{Time scales}

The autocorrelation functions contain time-scales which are related to the movement of the particle in the flow.  We identify two scenarios for the autocorrelations ($C_\text{mov}\kla{\tau}$  and $C_\text{tumb}\kla{\tau}$): as illustrated in \fig{fig:acMagn}, they are either conducting a weakly- or a critically-damped oscillation. With increasing turbulence level the oscillation is gradually changing towards the critically damped case and  for high propeller speeds ($\fimp>2.5\,\hertz$) no damped oscillation is observed (cf. Fig.~\ref{fig:acRot}). % independent of the flow type
  In order to extract meaningful time-scales we, therefore, fit two test-functions to each autocorrelation function. The functions are the transient solution of a weakly damped  harmonic oscillator:
\begin{equation}\begin{split}
f_w(\tau)=a_0\,\exp\kla{-\tau/\tau_\text{corr}}\cdot \sin\kla{2\pi f_\text{osc}+\phi_0},
\end{split}\end{equation}
and of a critically damped one:
\begin{equation}
f_d(\tau)=\exp\kla{-\tau/\tcorr}\cdot \kla{a_0+a_1\,\tau}.
\end{equation} 
$\tcorr, f_\text{osc}$ and $a_0, a_1, \phi_0$ are  fit-parameters. We return  $\tau_\text{corr}$ and (if available) $f_\text{osc}$  from the test-function which performs better in  approximating the autocorrelation. 
$C_\text{mov}\kla{\tau}$  and $C_\text{tumb}\kla{\tau}$ access motion and tumbling of the particle, respectively, and yield thus different time scales. For the oscillatory case,  $f_\text{osc}$ contributes additional details on the particle's motion.

\fig{fig:timesJim}  shows \tcorr~as a function of the impeller frequency and driving. We find that both  rotation-invariant  (\gleich{eq:ACjim1})  and rotating-sensitive (\gleich{eq:ACJim2}) function find very similar correlation times in the counter-rotating configuration. Moreover, \tcorr~of the particle   follows roughly a $\fimp^{-3/2}$ power-law as suggested by the scaling of the Kolmogorov time scale (it is $\tau_\eta\propto\eps^{-1/2}$ and $\eps \propto \fimp^3$). 
Furthermore, $\tcorr$ obtained from the rotation-sensitive function is independent of the way we drive the flow.
In contrast thereto, the rotation-invariant function gives  correlation times, $\tcorr$, which are larger and only little dependent on the impeller frequency if the impellers are co-rotating. 
That means one can distinguish co- from counter-rotating forcing by comparing the timescales of the two auto-correlation functions. 

Concerning the oscillation frequency $f_\text{osc}$ (not shown in figure), we find that the  rotation-sensitive autocorrelation senses the tumbling/wobbling of the particle, which is directly related to the particle's imbalance and  independent of the flow. The rotation-invariant autocorrelation on the other hand shows an oscillation frequency following the impeller frequency with $f_\text{osc}\sim\frac{2}{3}\fimp$.

%Our approach is related to the simple methods of either measuring the slope at $\tau=0$ or the time a certain value is crossed. %If  $f_{osc}$ is available we can estienables us to estimate  the tumbling frequency of the particle.
%We also tested the over-damped case (which is a sum of exponential decays), but it showed to be not numerically   robust. One can further determine the slope of $A(\tau)$ near $\tau=0$ \emph{i.e.}~the derivative at $\tau\approx 0$. 
%For  the strongly-damped case this corresponds to fitting  an exponential decay $f_e(\tau)=\exp\kla{-\tau/\tau_\text{corr}}$ to $A\kla{\tau}$. However, in the case of an oscillating  $A\kla{\tau}$ one obtains a value  proportional to the frequency of the oscillation but not to de-correlation. \\

\section{Discussion}
\label{sec:discussion}
After  briefly presenting the working principle of an instrumented particle measuring Lagrangian accelerations, we established a mathematical framework based on statistical moments and auto-correlation functions to analyze turbulent flows from the particle's signals.
In particular, we developed methods which are either invariant or sensitive to the rotation of the particle and its sensor in the flow. These methods perform well within the wide range of tested turbulence levels. With a smartPART one gets access to  correlation time scales of the flow, as well as the variance and flatness of the (translational)  acceleration. Comparing the rotation-sensitive and the rotation-invariant autocorrelation allows  distinguishing between different flow regimes, notably detecting long-time correlated large vortex structures as shown here with the co-rotating forcing of a \karman flow.   
In contrast to particle tracking methods the instrument particle returns one long trajectory instead of many short realizations. To that extent, it has to be noted that we limited our analysis  to the extraction of global flow features. In order to follow the evolution of  a slowly changing flow in time, these methods can, however, be extended to sliding windows. 
Work on adaptive filtering techniques is ongoing, in particular we are testing  the \emph{Empirical Mode Decomposition}, which  might be able to  separate the different contributions of the signal and thereby get even deeper insight into the flow.

% in contrast to tracer particles one
We emphasize that after usage  the particle can be easily extracted from the flow and then be reused and that by virtue of the developed mathematical framework no optical access is needed. The instrumented particle can therefore shed some light into flows that are not or hardly accessible up to now. 
Due to its continuous transmission  one flow configuration can be characterized within $\sim30\,\minute$.  
This technique is an  interesting tool for a fast quantification of a wide range of  flows  as they are found both in research labs and industry.

%\NOTE{Reason why we do not use Gyroscopes: electronic limitations but perhaps in the future if these limitations are met.}
%\NOTE{Alex made a remark on how buoyancy is handled: we should stress more that this is incorporated in $\atrans $, however this should only work for round smartPARTs }

\begin{acknowledgments}
This work was  supported by ANR-07-BLAN-0155.  
The authors want to acknowledge the technical help of Marius Tanase and Arnaud Rabilloud for the electronics, and of all the ENS machine shop staff. 
The authors also thank Michel Vo{\ss}kuhle, Micka\"el Bourgoin and Alain Pumir for many fruitful discussions.
\end{acknowledgments}

\bibliography{biblio}

\end{document}